\documentclass[%
reprint,
 amsmath,amssymb,
 aps,
 showkeys,
]{revtex4-2}

\usepackage{graphicx}% Include figure files
\usepackage{dcolumn}% Align table columns on decimal point
\usepackage{bm}% bold math
\usepackage{multirow}
\begin{document}

\preprint{APS/123-QED}

\title{Mie-scattering controlled all-dielectric resonator-antenna for bright and directional point dipole emission}% Force line breaks with \\
%\thanks{A footnote to the article title}%

%\collaboration{MUSO Collaboration}%\noaffiliation

\author{Mohammed Ashahar Ahamad}
 %\homepage{http://www.Second.institution.edu/~Charlie.Author}
%\email{ashaharamu2014@gmail.com}

\affiliation{%
 Department of Physics, Aligarh Muslim University, Aligarh, Uttar Pradesh %202002, India\\
}%

\author{Sughra Shaikh}
 %\homepage{http://www.Second.institution.edu/~Charlie.Author}
%\email{sughra.physics@gmail.com}
%\author{Nadeem Ahmed}%
% \email{khannadeemamu@gmail.com}
%\affiliation{%
% Department of Physics, Aligarh Muslim University, Aligarh, Uttar Pradesh %202002, India\\
%}%

\author{Faraz Ahmed Inam}
% \altaffiliation{Department of Physics, Aligarh Muslim University, Aligarh, U.P., India} %Lines break automatically or can be forced with \\
\email{faraz.inam.phy@amu.ac.in}
%\author{Nadeem Ahmed}%
% \email{khannadeemamu@gmail.com}
\affiliation{%
 Department of Physics, Aligarh Muslim University, Aligarh, Uttar Pradesh %202002, India\\
}%

%\affiliation{
% Third institution, the second for Charlie Author
%}%
%\author{Delta Author}
%\affiliation{%
% Authors' institution and/or address\\
% This line break forced with \textbackslash\textbackslash
%}%

%\collaboration{CLEO Collaboration}%\noaffiliation

\date{\today}% It is always \today, today,
             %  but any date may be explicitly specified

\begin{abstract}
Designing a deterministic, bright, robust, room temperature stable, on-demand solid-state single photon source has been a major demand in the field of quantum-photonics. For this, various single-photon resonator and antenna schemes are being actively explored. Here, using the Cartesian multi-polar decomposition of the excited Mie-scattering moments, we present the design of a all-dielectric coupled-dipolar antenna comprising of two dielectric (Tin-oxide, TiO$_2$) cylinders sandwiching a nanodiamond based nitrogen-vacancy (NV$^-$) center trapped in a poly-vinyl alcohol (PVA) matrix. The Mie-scattering resonant cavity formed in the middle PVA layer provides more than an order of magnitude decay rate or Purcell enhancement. The balancing of the electric and magnetic dipolar moments (a phenomenon commonly known as the Kerker condition) of the coupled TiO$_2$ cylinders under NV$^-$ dipole excitation, provides significant directionality to the radiation pattern. Using a collection lens with a numerical aperture (NA) of 0.9 the vertical collection efficiency (VCE) was observed to be around 80\% at the NV$^-$ center's zero-phonon line wavelength.    
\end{abstract}

\keywords{Cartesian multi-polar decomposition, Mie-scattering moments, Kerker condition, directional antenna, single-photon source and diamond NV center}%Use showkeys class option if keyword
                              %display desired
\maketitle

%\tableofcontents

\section{\label{sec:level1}Introduction:\protect}

The emerging quantum optical technologies like quantum key distribution, quantum information processing, quantum teleportation require on-demand, bright single-photon sources (SPS) \cite{Aharonovich2016, Atature2018a}. The SPS should be deterministic, robust, room-temperature operable, as well as indistinguishable. Many solid-state quantum emitters have emerged as prominent SPS due to their room-temperature scalability, photo-stability, and high quantum efficiency  \cite{Atature2018a, Buckley2012, Aharonovich2011, Lohrmann2017, Tran2015}. For deterministic or on-demand SPS, the quantum efficiency (QE) for photon emission should be close to unity, \textit{i.e.} whenever the source is triggered, a photon is generated. For applications in linear quantum computing, a theoretical study has set the minimum threshold CE value to be 2/3 (\~67\%) of the total emission\cite{Varnava2008}. Therefore, an ultra-bright solid-state SPS together with a CE value higher than 2/3 is required for implementing sustained quantum technological application.  

To increase the photon generation rate from deterministic SPS based on solid-state quantum emitters, various resonator schemes based on photonic cavities \cite{Indukuri2019, Zhang2018}, plasmonic nano-resonators \cite{Kleemann2017, Bogdanov2018} or hyperbolic meta-materials \cite{Kala2020, Galfsky2015} are actively explored. The plasmonic resonators have reported the maximum emission rate enhancement at room temperatures \cite{Koenderink2017, Akselrod2014, Bogdanov2018}. However, the inherent plasmonic losses renders the photon collection efficiency (CE) from these resonators. This renders the deterministic nature of these SPS and restricts it's application in quantum technologies. The CE is defined as \cite{Karamlou2018}: $ CE =\frac{1}{2Z_0}\frac{\int \int |E(\theta, \phi)|^2sin(\theta)d\theta d\phi}{P_{out}}$, with $P_{out}$ being the total power radiated by the dipole, $Z_0$ is the impedance of the surrounding medium, $|E|^2$ is the intensity of the electric-field in the far-field, $0\leq\phi\leq2\pi$ (the whole azimuth) and $0\leq\theta\leq sin^{-1}(NA)$ corresponds to all the angle within the numerical aperture (NA) of the collection lens.

%For  nanodiamond nitrogen-vacancy (NV$^-$) center, when the emission was coupled to a plasmonic patch antenna cum resonator, the maximum observed count rate was close to 20 Mcps for the case of single emitters  ***

%he point dipole is driven externally, \textit{e.g.}, using an excitation laser, at the emission frequency of the quantum emitter. This produces an emission pattern, with the radiated electromagnetic field acting back on the oscillating dipole emitter, known as the \textit{radiation reaction}. The emitted radiated field when acts back on the dipole emitter, does work on it and governs the complete radiation process of the dipole emitter. The \textit{radiation reaction} can be thought of as a process to pull energy out of the oscillating dipole and putting it into the available modes of the electromagnetic field. Larger will be the number of available modes, larger will be the power pulled out from the oscillating dipole****.

In classical electrodynamics, the emission rate from a single quantum emitter can be considered in terms of a radiating point dipole \cite{Xu1999, Barnes2020}. The emission radiation pattern is determined by the available density of photon states at the emission frequency. These photon states at the emission frequency are summed over the available wave-vectors and polarization, referred as the local density of optical states (LDOS) \cite{Novotny2012}. In terms of these electromagnetic modes, the LDOS, $\rho(\omega,r)$ for dipole emission is defined as \cite{Novotny2012, Koenderink2006}:
\begin{equation}
   \rho(\omega,r)=\sum_{k,\sigma}|\hat{d}\cdot{\Vec{E}_{k,\sigma}(r)}|^2\delta(\omega-\omega_{k,\sigma}). 
\label{LDOS}
\end{equation}
Here, $\hat{d}$ is the unit vector specifying the direction of the transition dipole moment, $\omega$ is the transition frequency, $\Vec{E}$ is the field at the source position, and the summation is over all wave-vectors ($k$) and polarization ($\sigma$). The value of $\Vec{E}(r)$ is the result of the superposition between the fields radiated directly by the dipole source and the fields scattered or reflected back from the surrounding structures and interfaces. The LDOS governs the complete radiation process of a dipole emitter and hence for any reliable control on the emission processes, the deterministic tuning of the LDOS is an essential requirement. 

\begin{figure*}
\centering
\includegraphics[width=5in]{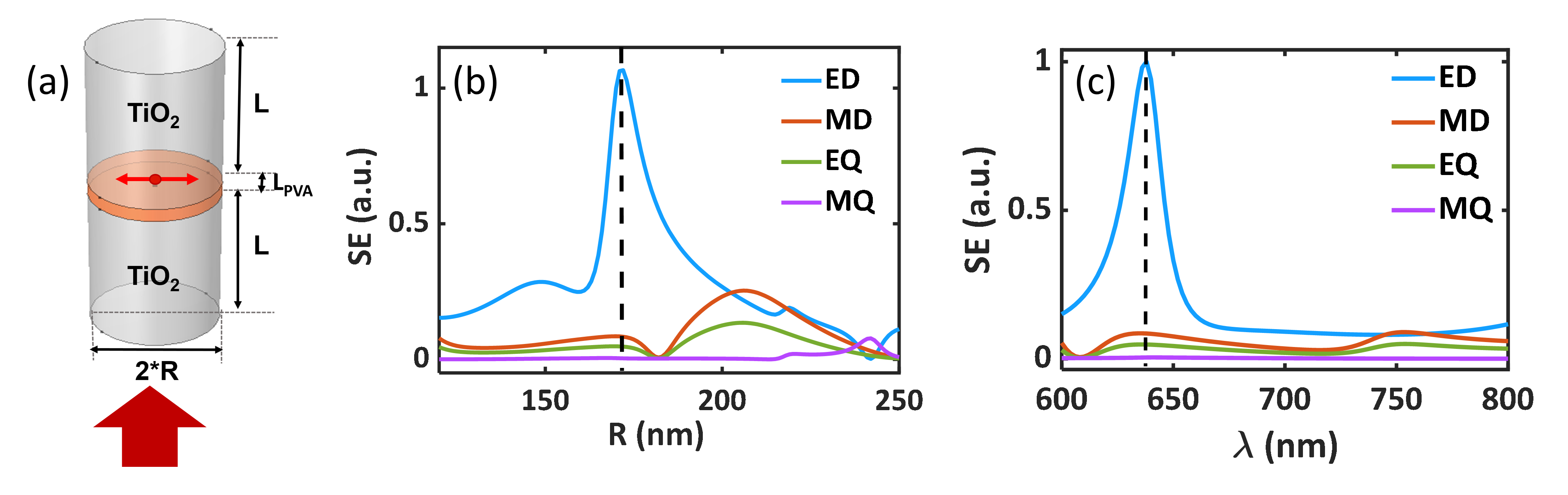}
\caption{(a) The schematic of the coupled-dipolar resonator antenna under plane-wave excitation from the bottom direction. (b) The radial optimization of the structure. The SE of the sandwiched PVA layer as a function of the structure radius. (c) The calculated SE spectrum of the PVA layer.}
\label{Schematic_cavity}
\end{figure*}

When electromagnetic fields are incident on a dielectric resonator with dimensions of the order of the excitation wavelength, various charge and current distributions are induced inside it. These will lead to the excitation of various multi-polar moments like the electric dipole (ED), magnetic dipole (MD), electric quadrupole (EQ), magnetic quadrupole (MQ), and higher order moments, called the Mie-scattering modes\cite{Kerker1969, Bohren1998}. The Mie-scattering resonances can control the field, $\Vec{E}(r)$ at the dipole sources position, therefore playing an important role in tuning the LDOS and in-turn the radiation process of a dipole emitter placed inside or in the vicinity of a resonator \cite{Alaee2018, Koshelev2020, Barnes2020}. The coherent superposition of the Mie-scattering moments in engineered photonic structures paves the way for directional scattering \cite{Koshelev2020, Shamkhi2019a}, high-Q supercavity modes in dielectric resonators \cite{Rybin2017}, bound states in continuum (BIC) in dielectric nanoparticle based metasurfaces\cite{Liu2019} or LDOS enhancement in silicon nanodisk based metasurfaces \cite{Khokhar2022}.

Here, we study the role of electromagnetic Mie-scattering modes of an all-dielectric coupled-dipolar antenna (Fig.~\ref{Schematic_cavity}a) in controlling the spontaneous emission rates as well as the radiation directionality of the NV$^-$ dipole emitter embedded inside it. The coupled-dipolar antenna comprises of two dielectric (Tin-oxide, TiO$_2$) cylinders sandwiching a poly-vinyl alcohol (PVA) layer encapsulating the nanodiamond based nitrogen-vacancy (NV$^-$) center. A Mie-scattering resonant cavity was observed to be formed in the middle PVA layer and it provides more than an order of magnitude decay rate or Purcell enhancement. Under, the NV$^-$ center dipole excitation, the balancing of the electric and magnetic dipolar Mie-scattering moments (a phenomenon commonly known as the Kerker condition) of the coupled TiO$_2$ cylinders, provides significant directionality to the radiation pattern. Using a collection lens with a numerical aperture (NA) of 0.9, the vertical collection efficiency (combining the collection from the top and bottom sides) was observed to be around 80\% at the Kerker condition ($\lambda = 647$ nm). With a bottom Ag reflector layer, the top collection was observed to reach about 70\%.

\section{\label{sec:level1}Results and discussion\protect}

\begin{figure*}
\centering
\includegraphics[width=5in]{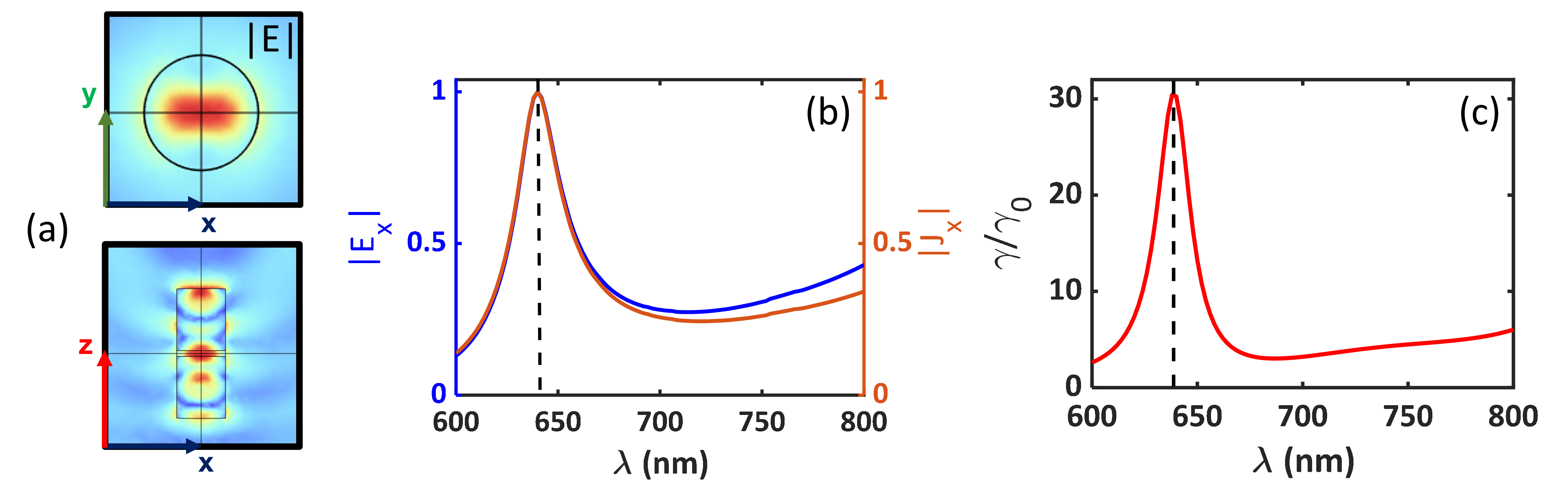}
\caption{(a) 2D-maps of the electric field profile within the structure under the plane-wave excitation (with $k$ vector along the z- or vertical axis and electric field $E$ along the x- direction). (b) The spectrum of the x-component of the electric field, $E_x$ and the current density $J_x$ at the center of the PVA cavity. (c) The computationally obtained decay rate spectrum of the nanodiamond NV$^-$ center placed at the center of the PVA cavity relative to a nanodiamond in air (Purcell enhancement).}
\label{Decay_rate}
\end{figure*}

The scattering efficiency (SE) corresponding to the individual Mie scattering moments of the sandwiched PVA layer were calculated analytically through the multi-pole decomposition technique \cite{Alaee2018}, using the electric field, $\textbf{E}(\textbf{r},\omega)$ and the current density $\textbf{J}(\textbf{r},\omega)$ profiles (obtained computationally using finite-element method (FEM) based commercial Comsol Multiphysics RF module) of the structure under longitudinal ($k$ vector along the z- or vertical axis) plane-wave excitation. The sandwiched PVA layer was observed to act as a planar cylindrical cavity. The cavity parameters (radius, $R$ and length, $L$) are optimized for NV$^-$ center's zero-phonon line (ZPL) at 638 nm \cite{Inam2013} to get a peak in the SE curve. Fig.~\ref{Schematic_cavity}b shows this optimization for the cavity radius, $R$. The optimized $R$ and $L$ were found to be 170 nm and 430 nm, respectively. Fig.~\ref{Schematic_cavity}c shows the calculated SE spectrum of the PVA cavity. The electric dipole (ED) moment shows a dominant resonant peak at the NV$^-$ center's ZPL at 638 nm. The strength of the peak was observed to depend on the sandwiched PVA layer thickness. Considering the nanodiamond size to be about $30-40$ nm, the PVA layer thickness is taken to be 40 nm here. 

2D-maps of the electric field profile within the structure under the plane-wave excitation (with $k$ vector along the z- or vertical axis and electric field $E$ along the x- direction) is shown in Fig.~\ref{Decay_rate}a. The electric field is observed to be confined within the PVA domain showing that the PVA layer is acting as a cavity, enhancing the fields within. Fig.~\ref{Decay_rate}b shows the spectrum of the x-component of the electric field, $E_x$ and the current density $J_x$ at the cavity center. Both $E_x$ and $J_x$ shows a resonance at the NV$^-$ center's ZPL wavelength 638 nm. The field confinement and enhancement in the cavity will consequently enhance the LDOS for a dipole emitter placed within the PVA cavity. For a horizontally oriented (in-plane) point dipole emitter located at the center (Fig.~\ref{Schematic_cavity}a), the radiated power can be calculated in terms of the average work-done per oscillation
cycle $W(t) = \langle J(t) \cdot E(t)\rangle$ or by
calculating the power flux passing through a closed surface enclosing the dipole source \cite{Xu1999}. The computationally obtained decay rate spectrum of the nanodiamond NV$^-$ center placed at the center of the PVA cavity relative to a nanodiamond in air (Purcell enhancement) was observed to follow the pattern of $\langle J(t) \cdot E(t)\rangle$ (Fig.~\ref{Decay_rate}c).

\begin{figure*}
\centering
\includegraphics[width=5in]{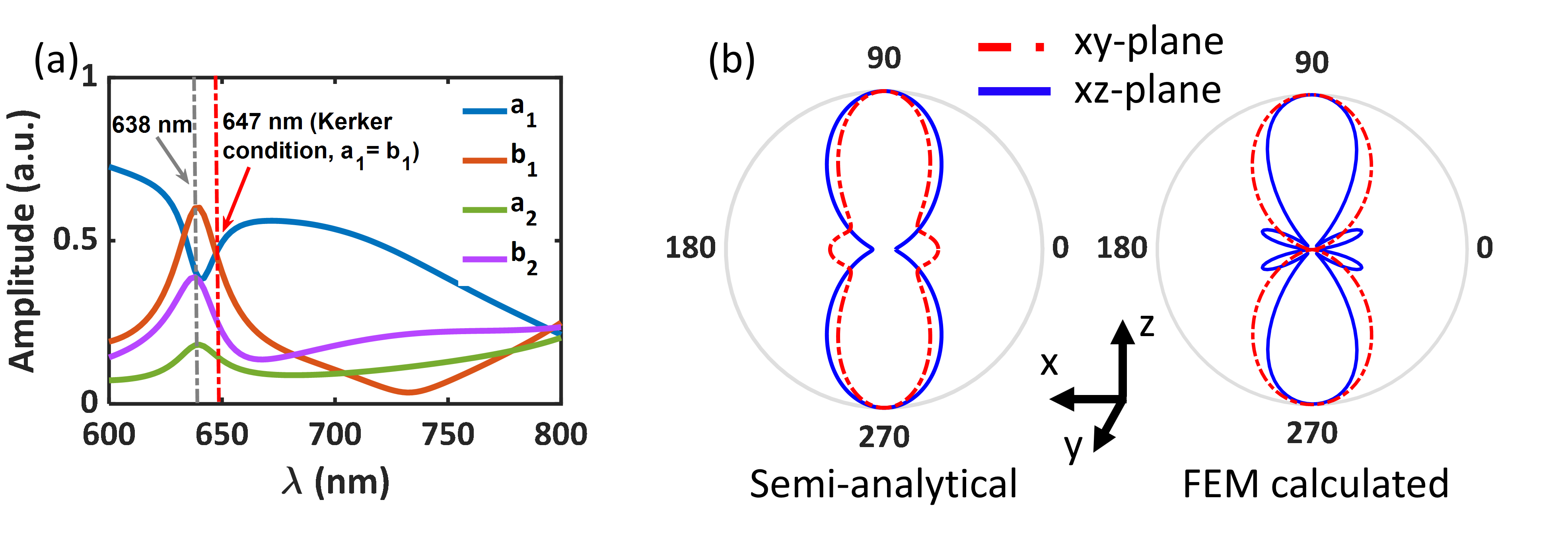}
\caption{(a) The spectrum of the induced multi-polar moments (electric dipole, ED (a$_1$), magnetic dipole, MD (b$_1$), electric-quadruple, EQ (a$_2$) and magnetic quadruple, MQ (b$_2$)) of the TiO$_2$ cylinder under embedded dipole excitation. (b) The semi-analytically (Cartesian multi-polar decomposition based) as well as computationally (FEM based) calculated radiation patterns for NV$^-$ center dipole emission at 638 nm (ZPL).}
\label{SE_dipole_and_radiation_pattern}
\end{figure*}  

We now study the influence of the whole coupled dipolar TiO$_2$ antenna (Fig.~\ref{Schematic_cavity}a) on the radiation pattern of the embedded NV$^-$ center dipole emitter. For a dipole emitter, radiating very close to an antenna, the emitter's radiation will induce various electromagnetic
multi-polar moments (Mie-scattering moments) within the antenna. These excited moments will also radiate electromagnetic wave into the far-field and these radiations will interfere with the radiation emitted by the dipole emitter \cite{Jackson1975, Qin2022}. To analyse this effect, we start with the Cartesian multi-polar decomposition of the excited Mie-scattering moments of the dipolar TiO$_2$ antenna under the embedded dipole excitation \cite{Alaee2018}. Fig.~\ref{SE_dipole_and_radiation_pattern}a shows the spectrum of the induced multi-polar moments (electric dipole (a$_1$), magnetic dipole (b$_1$), electric-quadrupole (a$_2$) and magnetic quadrupole (b$_2$)) of the TiO$_2$ cylinder under embedded dipole excitation. The far-field differential scattering cross-section of the multi-polar Mie-scattering moments is written as \cite{Bohren1998, Miroshnichenko2018}:  
\begin{equation}
  \frac{d\sigma^{scatt}}{d\Omega}=\frac{1}{k_0^2}\left[cos^2\phi |S_{||}(\theta)|^2 + sin^2\phi |S_{\perp}(\theta)|^2\right], 
\label{Scattering_distribution}
\end{equation}
with $S_{\perp}(\theta)$ and $S_{||}(\theta)$ are the polarized scattering waves perpendicular and parallel to the scattering plane, which are written as:
\begin{subequations}
\begin{flalign}
    S_{||}(\theta)=\Sigma_{n=1}^{n=\infty}\frac{2n+1}{n(n+1)}\left[a_n\frac{dP_n^{(1)}(cos\theta)}{d\theta} + b_n\frac{P_n^{(1)}(cos\theta)}{sin\theta}\right],\\
    S_{\perp}(\theta)=\Sigma_{n=1}^{n=\infty}\frac{2n+1}{n(n+1)}\left[b_n\frac{dP_n^{(1)}(cos\theta)}{d\theta} + a_n\frac{P_n^{(1)}(cos\theta)}{sin\theta}\right].
\end{flalign}
\label{Scattering_distribution}
\end{subequations}

\begin{figure*}
%\centering
\includegraphics[width=3.5in]{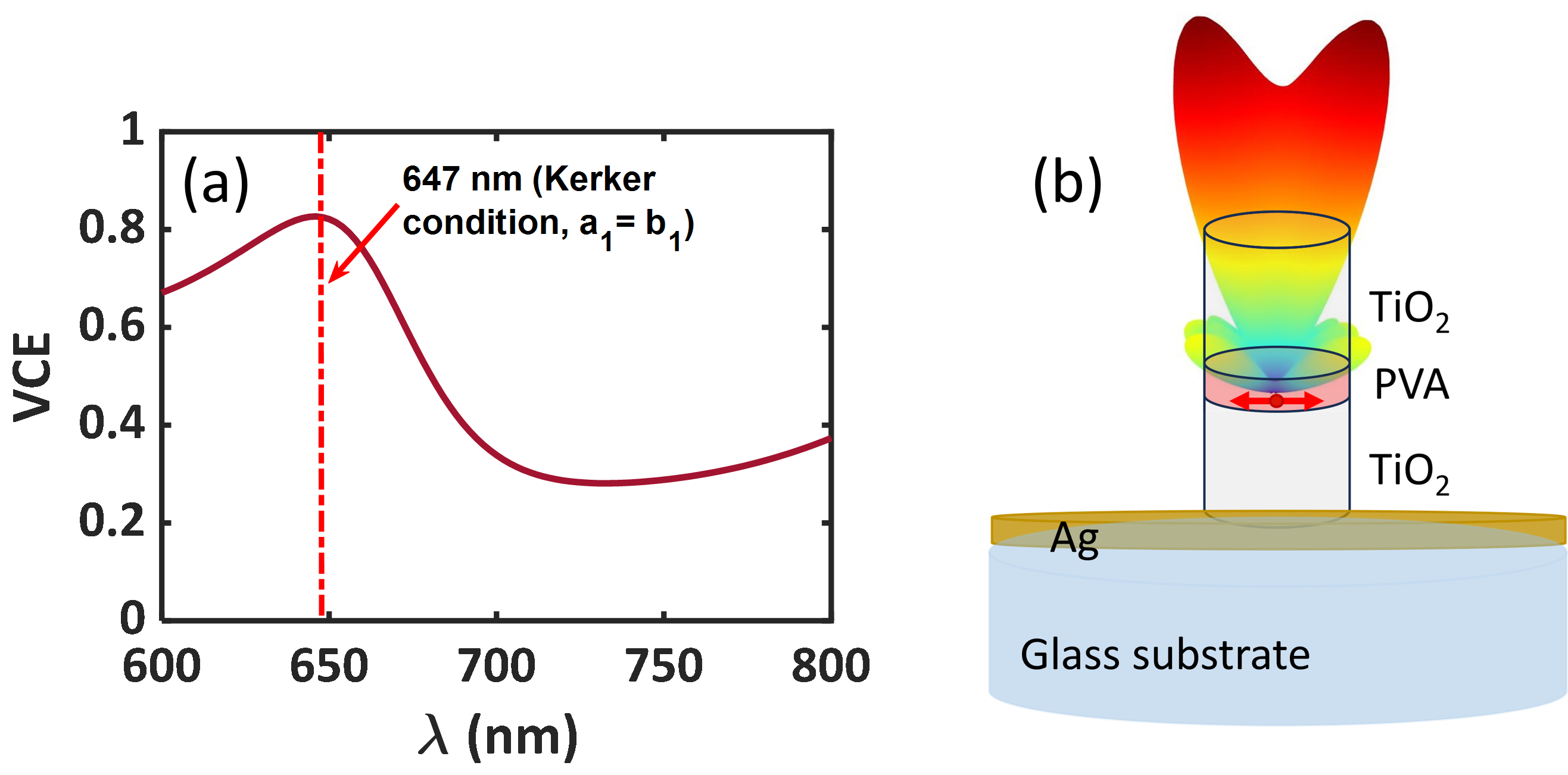}
\caption{(a) The VCE spectrum of the embedded NV$^-$ center based point dipole emitter in the coupled-dipolar TiO$_2$ antenna. (b) The 3-D radiation pattern of the embedded dipole emitter at the Kerker point ($\lambda = 647$ nm) with a Ag bottom reflector layer.}
\label{CE_and_radiation_pattern}
\end{figure*}  

Here, $\theta$ and $\phi$ are the polar and azimuthal angles. $\theta = 0$ will be the forward direction and $\theta = \pi$ will be the backward direction. $\phi = 0$ corresponds to the in-plane and $\phi = \pi/2$ corresponds to the out-of-plane directions. All the individual multi-polar moments will act as a source of electromagnetic radiation/waves. The physical picture of the dipole emitters radiation pattern will be a superposition of all these waves and is given as \cite{Jackson1975, Qin2022}: i) destructive interference between the excitation electric dipole moment, p$_0$ and the induced electric dipole moment of the antenna, a$_1$; and ii) superposition of all the other excited multi-polar moments. Fig.~\ref{SE_dipole_and_radiation_pattern}b shows the semi-analytically (Cartesian multi-polar decomposition based) as well as computationally (FEM based) calculated radiation patterns for NV$^-$ center dipole emission at 638 nm (ZPL). Since, there is a TiO$_2$ cylinder along both the top and bottom directions of the embedded dipole emitter (oriented along the x-axis), the radiation is symmetric along both the directions.

Using a collection lens with a numerical aperture (NA) of 0.9 \cite{Kala2020}, the vertical collection efficiency, VCE (combining the collection from the top and bottom sides) spectrum is shown in Fig.~\ref{CE_and_radiation_pattern}a. The VCE shows a peak 647 nm, where the ED (a$_1$) = MD (b$_1$) (first Kerker condition \cite{Shamkhi2019a}) and the contributions from the other higher order moments is small. The Kerker effect is known to provide forward or backward directionality to the scattering pattern of the antenna \cite{Shamkhi2019a}. This results in a high VCE value of 80\%. At the NV$^-$ center ZPL wavelength (638 nm), the VCE is about 75\%. In-order to maximize the collection from the top side, a prospective design is shown in Fig.~\ref{CE_and_radiation_pattern}b. It uses a silver (Ag) based reflector layer along the bottom, below the TiO$_2$ cylinder. In this scheme, the collection efficiency was observed to be about 70\% at the Kerker point (647 nm).

\section{\label{sec:level1}Conclusions:\protect}
Electromagnetic multi-polar Mie-scattering moments of an antenna plays a role in tuning it’s far-field radiation pattern. Here, in this study we showed how the Mie-scattering moments of a cavity cum antenna, helps in controlling the complete spontaneous emission process of a point dipole emitter, the nanodiamond NV$^-$ center based SPS here. The Mie-resonance of the ED moment of the planar-cylindrical cavity formed in the sandwiched PVA layer results in significant field confinement and enhances the decay-rate of the embedded NV$^-$ center based point dipole emitter by more than an order of magnitude. Further, the Cartesian multi-polar decomposition of the Mie-scattering moments of the top and bottom TiO$_2$ cylinders under the point dipole excitation, were observed to control the radiation pattern of the embedded dipole emitter. The radiation directionality along the vertical top and bottom directions were found to be maximum at the Kerker point ($\lambda = 647$ nm). At the NV$^-$ center's ZPL wavelength, the VCE was observed to be about 80\%. With a bottom Ag reflector layer, the top collection was observed to reach about 70\%. All these CE values are well above the 2/3 threshold required for practical applications in linear quantum computing \cite{Varnava2008}. 

Combining the decay-rate enhancement and the CE values, we expect the photon collection rates from an embedded NV$^-$ center is expected to be few 100 MHz \cite{Inam2023}. Our results are therefore substantial to accelerate the research for generating on-demand SPS for quantum photonic applications. Further, the understanding of the tuning of the dipole-emitter's radiation pattern with the excited multi-polar Mie-scattering moments of the antenna will help in controlling the emission coupling between the individual dipole emitters and hence the dependent phenomenons like super-radiance, many-body interactions, etc.   

\begin{acknowledgments}
The authors would like to acknowledge the financial support from the Department of Science and Technology (DST), India for funding under the FIST Program 2020 and Core Research Grant (CRG/2021/001167). 
\end{acknowledgments}

\bibliography{MyCollection.bib}% Produces the bibliography via BibTeX.

\end{document}